# Graphene Based Adaptive Thermal Camouflage

Omer Salihoglu[1], Hasan Burkay Uzlu[1], Ozan Yakar[1], Shahnaz Aas[1],

Osman Balci[1], Nurbek Kakevov[1], Sinan Balci[2], Selim Olcum[3], Sefik Süzer[4], and Coskun Kocabas[1,5]

[1]Bilkent University, Department of Physics, 06800, Ankara Turkey

[2]Department of Photonics, Izmir Institute of Technology, 35430 Izmir, Turkey

[3]Massachusetts Institute of Technology, Department of Biological Engineering

Cambridge MA 02139-4307,

[4]Bilkent University, Department of Chemistry, 06800, Ankara Turkey

[5]School of Materials and National Graphene Institute, University of Manchester, Oxford Rd, Manchester, M13 9PL, UK

Email: coskun.kocabas@manchester.ac.uk

ABSTRACT: In nature adaptive coloration has been effectively utilized for concealment and signaling. Various biological mechanisms have evolved that can tune the reflectivity for visible and ultraviolet light. These examples inspire many artificial systems for mimicking adaptive coloration to match the visual appearance to their surroundings. Thermal camouflage, however, has been an outstanding challenge which requires an ability to control of the emitted thermal radiation from the surface. Here we report a new class of active thermal surfaces capable of efficient real-time electrical-control of thermal emission over the full infrared (IR) spectrum without changing the temperature of the surface. Our approach relies on electro-modulation of IR absorptivity and emissivity of multilayer graphene via reversible intercalation of nonvolatile ionic liquids. The demonstrated devices are light (30 g/m$^2$), thin (<50 μm) and ultra-flexible which can conformably coat their environment. In addition, combining active thermal surfaces with a feedback mechanism, we demonstrate realization of an adaptive thermal camouflage system which can reconfigure its thermal appearance and blend itself with the thermal background in a few seconds. Furthermore, we show that these devices can disguise hot objects as cold and cold ones as hot in a thermal imaging system. We anticipate that, the electrical control of thermal radiation will impact on a variety of new technologies ranging from adaptive IR optics to heat management for outer space applications.

The ability to control thermal radiation from a hot object has both scientific[1-5] and technological importance[2, 6-8]. The radiated thermal energy per unit area from a hot surface is characterized by the Stefan–Boltzmann law, $P = \varepsilon \sigma T^4$ where $\varepsilon$ is the emissivity of the surface, $\sigma$ is the Stefan–Boltzmann constant and $T$ is the temperature of the surface. The emissivity is the only material-dependent parameter that varies with the wavelength and temperature. At thermodynamic equilibrium, Kirchhoff's radiation law connects the wavelength-specific thermal emissivity with the optical absorption of the surface as $\varepsilon(T,\lambda) = \alpha(T,\lambda)$. One can engineer the thermal radiation by coating the surface with photonic crystals[5, 9-11] or plasmonic structures[12]. The dynamic control of thermal radiation, however, requires ability to alter optical absorption via electrical means. Phase change materials[13-16], quantum wells[17], electrochromic dyes[18], ferroelectric materials[19] or plasmonic resonators[12, 20, 21] have all been investigated for tunable infrared emission. These research efforts on dynamic control of thermal radiation have encountered various problems such as, low tunability[19, 20, 22], narrow spectral window[17], slow response speed[18] and rigid substrates[17]. Electrochromic materials have been the most promising one[23-25], however, the requirement of a top metallic contact layer and volatile electrolytes limit their performance (see the benchmarking in Table S1). These challenges have been hindering the realization of adaptive thermal camouflage systems.

Graphene provides new perspectives to control electromagnetic radiation in a very broad spectral range, from visible to microwave frequencies[26-28]. Optical absorption of graphene can be tuned by electrostatic gating owing to the Pauli blocking[29,30]. Although, optical response of graphene has been studied extensively, the use of graphene for dynamic control of thermal radiation has remained unexplored because of the small optical absorption (< 2%) in mid-IR region[31]. In this work, we developed a new class of active thermal surfaces using multilayer graphene which yields significant tunable optical absorption in IR region. Since thermal radiation originates from the very top surface, top-gating or electrolyte gating schemes are not suitable for the control of thermal radiation. These gating methods generate either buried graphene surfaces or low electrostatic doping[29,32], which yields negligible IR modulation. None of the previously reported graphene devices by our group and others is suitable for dynamic control of thermal radiation. Therefore, we introduce a new gating scheme using an inverse device structure which leads intercalation of a nonvolatile ionic liquid into graphene layers from the porous substrate. This inverse device configuration yields an uncovered graphene surface with tunable charge density and Fermi energy. Figure 1a shows the schematic of the active thermal surface consisting of a multilayer-graphene electrode on a porous polyethylene (PE) membrane and a back gold-electrode. We synthesized multilayer-graphene on nickel foils using a chemical vapor deposition method and then transferred them on PE membrane which is IR transparent and can hold the electrolyte (room-temperature ionic liquid, RTIL).

The thermal radiation emitted by the device mainly originate from the top graphene electrode since the emissivity of gold-coated substrate is very low (<0.01) due to its highly reflective nature, and the PE membrane is IR transparent. Figure 1b illustrates the working principle of the active thermal surface. Under a voltage bias, the ionic liquid intercalates into the graphene layers, and dopes them. As a result of doping, the charge density on graphene increases and Fermi-level shifts to higher energies, which suppresses the IR absorption and thus the emissivity of the graphene electrode[30]. Figure 1c and 1d show the thermal camera images of the fabricated device at 0 and 3V, respectively. At 0 V, the temperature profile of the background (author's hand) can be seen through the device. However, at 3 V, the emissivity of the device is significantly suppressed, which screens the background temperature profile (Movie S1). The emissivity of the device can be switched between high and low states many times with response time< 1s. These devices are thin, light, and flexible that can easily cover their environment (Figs. S1 and S2).

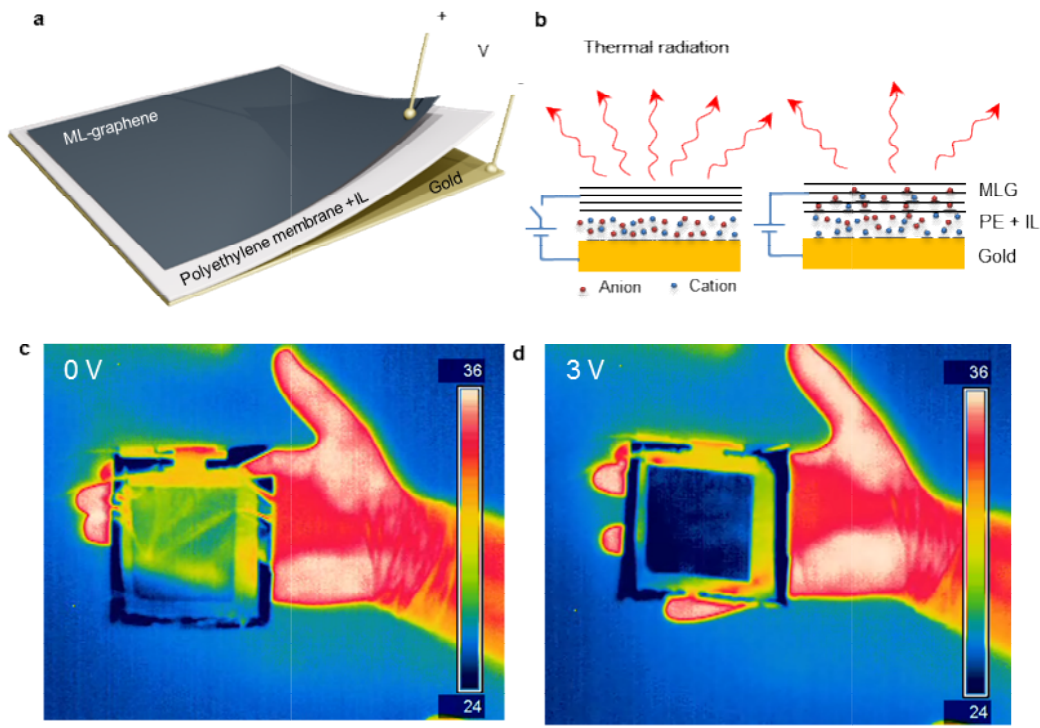

**Figure 1.** Active thermal surfaces. (a) Schematic drawing of the active thermal surface consisting of a multilayer-graphene electrode, a porous polyethylene membrane soaked with a RTIL and a back gold-electrode coated on heat resistive nylon. (b) Schematic representation of the working principle of the active thermal surface. The emissivity of the surface is suppressed by intercalation of anions into the graphene layers. (c-d) Thermal camera images of the device placed on the author's hand under the voltage bias of 0 and 3V, respectively.

To quantify the performance of the fabricated active thermal surfaces, we first placed them on a hot plate at 55 °C and recorded the thermal images (Figure 2a, 2b and Figure S1 and Movie 2), at different bias voltages between 0 and 4 V. Note that, the voltage range is limited by the electrochemical window of the room temperature ionic liquid[33]. We obtained the best performance with the IL [DEME][TFSI] which yields relatively large electrochemical window up to 4 V. The thermograms show substantial change in the thermal appearance which is quite homogenous over a large area device (10x9 cm$^2$). The IR camera renders the thermograms assuming a constant emissivity of 1. Although the temperature of the device is the same, the gold electrode appears cold at high voltages due to its low emissivity.

First, we measured the IR spectrum of the emitted radiation at different bias voltages (Figure 2c) using a Fourier transform infrared spectrometer (FTIR). The modulation of spectral radiance of the device covers the full mid-infrared range. The intensity of the spectrum decreases by a factor of 2.5 at 3.5 V over a broad range. To measure the variation of the total emitted thermal power from the device, we used a thermopile sensor which performs a differential measurement with respect to the room temperature (inset in Figure 2d). We recorded the output voltage of the sensor as we scanned the bias voltage between 0 and 4 V with a scan rate of 0.01 V/s (Supplementary Movie 2). To block the background radiation, we used a 3-inch-silicon-wafer coated with 100 nm thick gold film which has very low emissivity (<0.1). The voltage dependence of emitted power from the device is shown in Figure 2d. We observed a clear step-like behavior with a threshold voltage of 2 V. The emitted thermal power is reduced by a factor of 2.5 at a bias voltage of 3.5 V. These numbers agree very well with the spectral measurements.

To calculate the emissivity of device, we used a carbon nanotube forest as a reference black surface which has emissivity close to 1 (Figure S3)[34]. The extracted emissivity of the multilayer graphene at 10 μm is reduced from 0.76 down to 0.33 as we scanned the voltage from 0 to 3.5 V (Scattered plot in Figure 2d). Variation of the total radiated power and the extracted emissivity values show similar voltage dependence indicating that the variation of emissivity with the bias voltage is nearly constant over the mid-IR range. The intercalation process is reversible and the device can be switched between high and low emissivity values with a time constant of 0.5 s.

Our results suggest that, the observed suppression of the emissivity is due to the suppression of IR absorption of multilayer graphene via intercalation of ionic liquid. To further quantify the intercalation process, we measured variation of the sheet resistance of ML-graphene using four-point resistivity method (inset in Figure 2e). Similarly, the sheet resistance of the graphene electrode shows a step like variation from 33Ω down to 0.6 Ω (Figure 2e). The sheet resistance and the emissivity of ML-graphene are correlated. As the layer number increases both sheet resistance and emissivity decrease (Figure S4). To gain more insight about the mechanism behind the electrical control of thermal radiation, we performed *in situ* optical characterization of the ML-graphene electrodes (Figures S5 and S6). We observed that the transmittance of ML-graphene decreased substantially whereas the reflectivity increased due to the high level of doping. We also tested similar devices with single-layer graphene and observed slight modulation (<2% increase) of thermal radiation due to enhanced inter-band absorption (Figure S7). These results and our electromagnetic simulations reveal that, both inter-band and intra-band transitions of the ML-graphene contribute to the observed emissivity modulation in the IR spectrum[30, 35, 36] (Figure S8). The tunable high mobility free carriers on graphene layers are responsible for the control of the emissivity[37, 38].

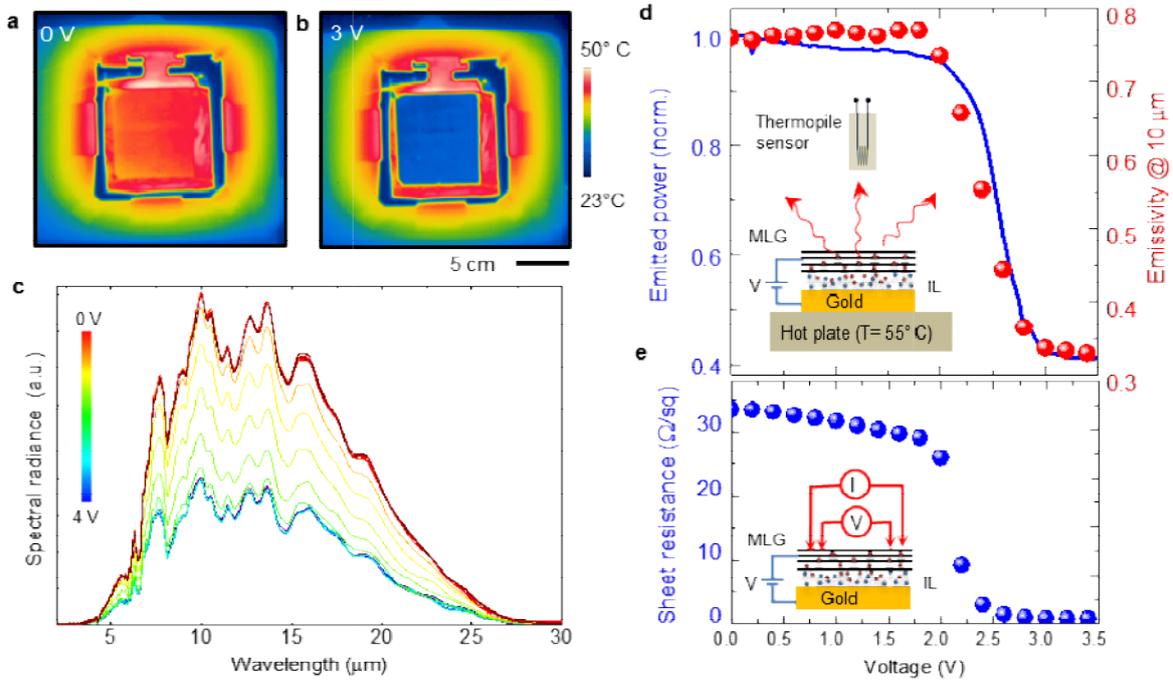

**Figure 2.** Voltage-controlled thermal emission. (a-b) Thermal camera images of the fabricated device biased at 0 V to 3 V, respectively. The device is placed on a hot plate kept at a temperature of 55° C. (c) Spectrum of the thermal radiation from the device at different bias voltages. (d) Voltage dependence of the emitted thermal power (blue line) and extracted emissivity (red scattered data) at the wavelength of 10 μm. The thermopile radiation sensor is placed 1 cm away from the device sitting on a hot plate. The emissivity is calculated using the carbon nanotube forest as a reference. The inset shows the experimental set-up used for measuring the voltage dependence of thermal radiation. (e) The sheet resistance of the multilayer graphene electrode plotted against the bias voltage. The inset shows the four-point measurement setup.

Using the nonvolatile RTIL electrolyte allows us to operate these devices also in ultrahigh vacuum conditions. This ability is critical for some applications such as active thermal shields for outer space applications[23], as well as utilization of surface characterization tools such as X-ray photoelectron spectroscopy (XPS), which can elucidate the operation of the devices in a chemically specific fashion. Although, intercalation of graphitic materials with metallic ions have been studied extensively[35], intercalation of ionic liquids remains relatively unexplored[33]. Our device lay-out (Figure 3a) provides a unique advantage to characterize the intercalation process. The ionic liquid contains two nitrogen atoms (Figure 3b); one with a positive (quaternized nitrogen) and the other with a negative (imide nitrogen) charge that yield well resolved two N1s peaks. Figure 3c shows the recorded C1s, N1s and F1s region of XP spectra at different bias voltages. These spectral evolutions provide a wealth of information about the operation of the device. The appearance of N1s and F1s peaks after 1.5V indicates the onset of the intercalation process and the threshold voltage. Since XPS probes the very top surface (~10 nm) appearance of F1s and N1s peaks shows that the ions can efficiently intercalate the thick active surface (>100 graphene layers). The intensity of C1s decreases with increasing voltage due to the partial coverage of the top surface with the IL. The C1s peak of the -$CF_3$ group associated with IL also appears after the threshold voltage. Although, the graphene surface is grounded, the binding energy of C1s also experiences a small shift with the applied bias, from 284.37 to 283.67 (Figure 3e) most likely due to the shift in the Fermi energy of graphene[39]. Interestingly, we observed co-intercalation of anions and cations of the ionic liquid with a significant charge imbalance > 20% (the ratio of $N^-$ to $N^+$). This charge imbalance (due to mobile and quasi-independent ions) is responsible for electrostatic doping on graphene layers. When we apply negative bias voltage, the charge imbalance is reversed (Figure 3d). Our results shows that intercalation of ionic liquid into multilayer graphene yields effectively a charge imbalance with a charge excess of about 1 ion for ~200 C atoms of the intercalated active layer (Figure S9). This direct observation of the chemical contents of intercalate with related electronic properties of the graphene layers will further guide us to optimize the device operation.

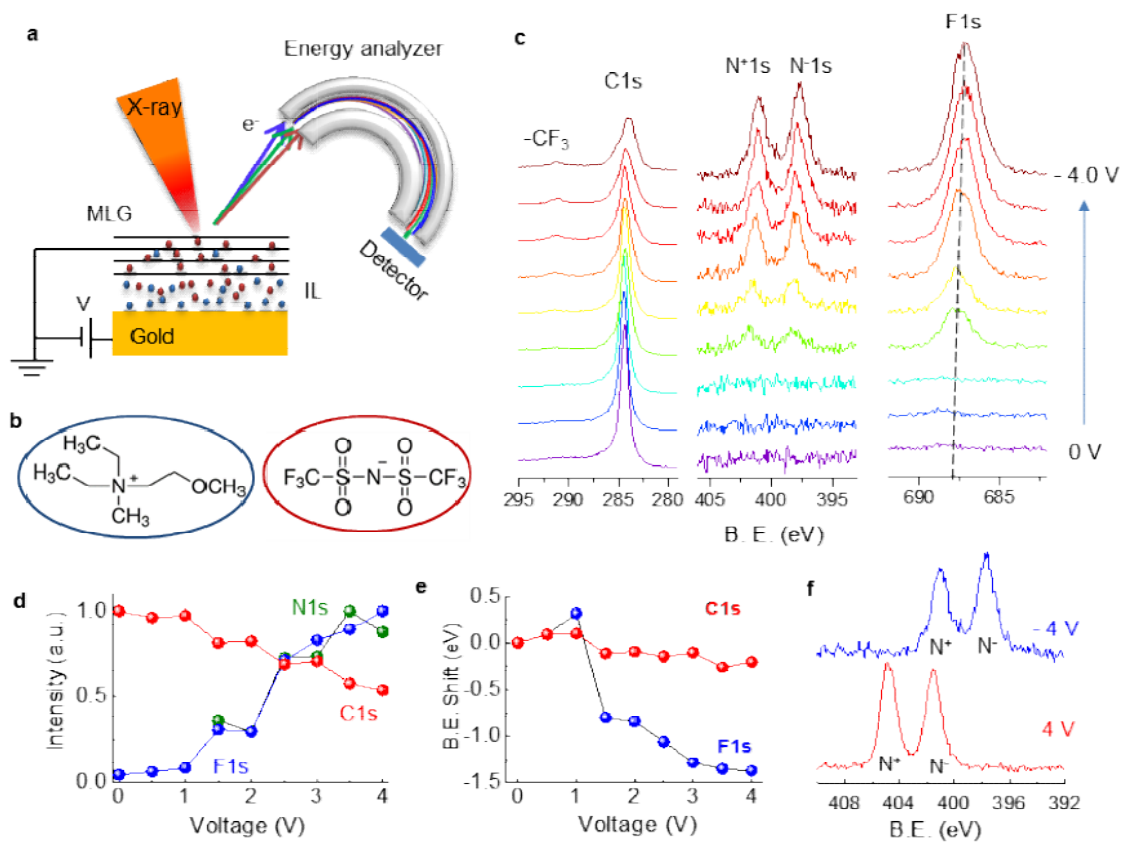

**Figure 3.** *In situ* XPS characterization of the active thermal surfaces. (a) Experimental setup used for the operando-XPS. (b) Chemical structure of ionic liquids. Positively and negatively charged nitrogen ions enable monitoring of the chemical content of intercalates. (c) XPS spectra recorded from the surface of device under bias voltages between 0 to 4V. The spectra were recorded in ultra-high vacuum $10^{-8}$ torr. (d) Variation of the normalized intensities and binding energy of C1s, N1s, and F1s. (e) The variation of the binding energy of C1s and F1s. (f) XPS spectra of N1s showing the charge imbalance for positive and negative bias voltages.

To show one promising application of the developed thermal surfaces, we now would like to demonstrate a functional adaptive camouflage system. In nature, animals developed adaptive camouflage techniques using specialized cells that enable active feedback mechanisms to adjust the skin color and texture[1, 40]. Our strategy uses thermal emission as a feedback. Figure 4a shows the working principle of the adaptive thermal camouflage system. The body temperature of the device is set to 40 °C. The thermocouple measures the actual surface temperature and sends the sensory information to the circuit which uses the thermal radiation from the device as a feedback and yield a control signal to adjust the thermal radiation. The algorithm minimizes the difference between the surface temperature and the apparent temperature of the device. Although the body temperature of the device is constant, by tuning the emissivity of the surface with the control voltage, this device can blend itself with the time varying thermal background. Figure 4b shows the varying surface temperature (red curve) and apparent temperature of the device (blue curve). After the optimization of the feedback gain, the apparent temperature follows the surface temperature with a small time delay of< 5 s (Figure S10). When we set a large gain in the control circuit, we observe large oscillations in the apparent temperature, but eventually, the apparent temperate reaches that of the background (Figure S11). This device can operate in the temperature range between 38 and 25 °C.

The dynamic range of the camouflage system depends on many factors; such as the body temperature of the device, modulation of the emissivity, the surface temperature, the background temperature (from the environment) and quality of thermal contact between the object and the active surface. To obtain the more insight for the operation range and further quantify the experimental observations, we developed a model for the apparent temperature. The thermal camera renders the temperature of a surface from the detected radiation which includes radiation from the surface and the reflected background (environment) radiation as $\varepsilon_c T_a^4 = \varepsilon_0 T_0^4 + R_0 T_b^4$ where $T_a$, $T_0$ and $T_b$ represent apparent, body and background temperatures, respectively and $\varepsilon_0$ is the emissivity of the surface and $\varepsilon_c$ is the emissivity used by the camera. We can write reflectivity of the surface as $R = 1 - A = 1 - \varepsilon_0$ where A is the absorption of the surface. (Note that the transmission of the device is 0 due to the gold electrode.) The solid lines in Figure 4c shows the relation between apparent temperature and the actual body temperature for different emissivity range from 0 to 1. For this calculation, we used background temperature of 26.7 °C. We first verify these calculations using a gold-coated surface ($\varepsilon_{Au}$~0) and carbon-nanotube sample ($\varepsilon_{CNT}$~1). Gold-coated surface always shows the background temperature due to the perfect IR reflectivity however, CNT sample shows the actual body temperature due to perfect emissivity (no reflectivity, see Figure S12). Apparent temperature of our device varies between these values depending on the emissivity ($\varepsilon$~ 0.3-0.8) and body temperature. Figure 4c reveals three intriguing results due to the interplay between the radiation and reflection. First, the dynamic range of the active surface increases with the temperature difference between the body and the background. Second, when the body temperature is the same with background, the apparent temperature of the device does not change with the voltage. The suppression of the emissivity is compensated by the increasing reflectivity. Third, when the body temperature is lower than the background, the apparent temperature increases with decreasing emissivity (increasing voltage). When the voltage was applied, the cold surface looks hotter. Therefore, the voltage controlled emissivity and reflectivity of ML-graphene enables us to design new camouflage systems that can disguise hot surfaces as cold and cold ones as hot in a thermal imaging system. When the surface is hotter than the background temperature, the thermal emission is dominant. Suppression of the emissivity of the surface yields colder appearance. However, when the object is colder than the background temperature, the reflection of the background radiation is dominant. Increasing concentration of high mobility carrier on the graphene surface under a bias voltage yields a hotter appearance in thermal imaging systems.

The thickness of the multilayer graphene is another important parameter that defines the modulation range of the emissivity. We fabricated and characterized a series of devices with varying the thickness of the active graphene layer. Figure 4d shows the variation of the measured and calculated emissivity with the layer number. The maximum emissivity of 0.8 can be obtained with 100 layers of graphene. Thicker or thinner films yield less emissivity due to larger reflectivity or smaller absorption, respectively. In Figure 4d, we also show the measured emissivity for the doped graphene (at 3.5V, blue dots). We observed that minimum emissivity also varies with the layer number which is likely due to inefficient intercalation for thick films and residual infrared absorption of doped graphene in Pauli blocking regime which is not fully understood yet. The maximum emissivity modulation can be obtained with 150 layers of graphene.

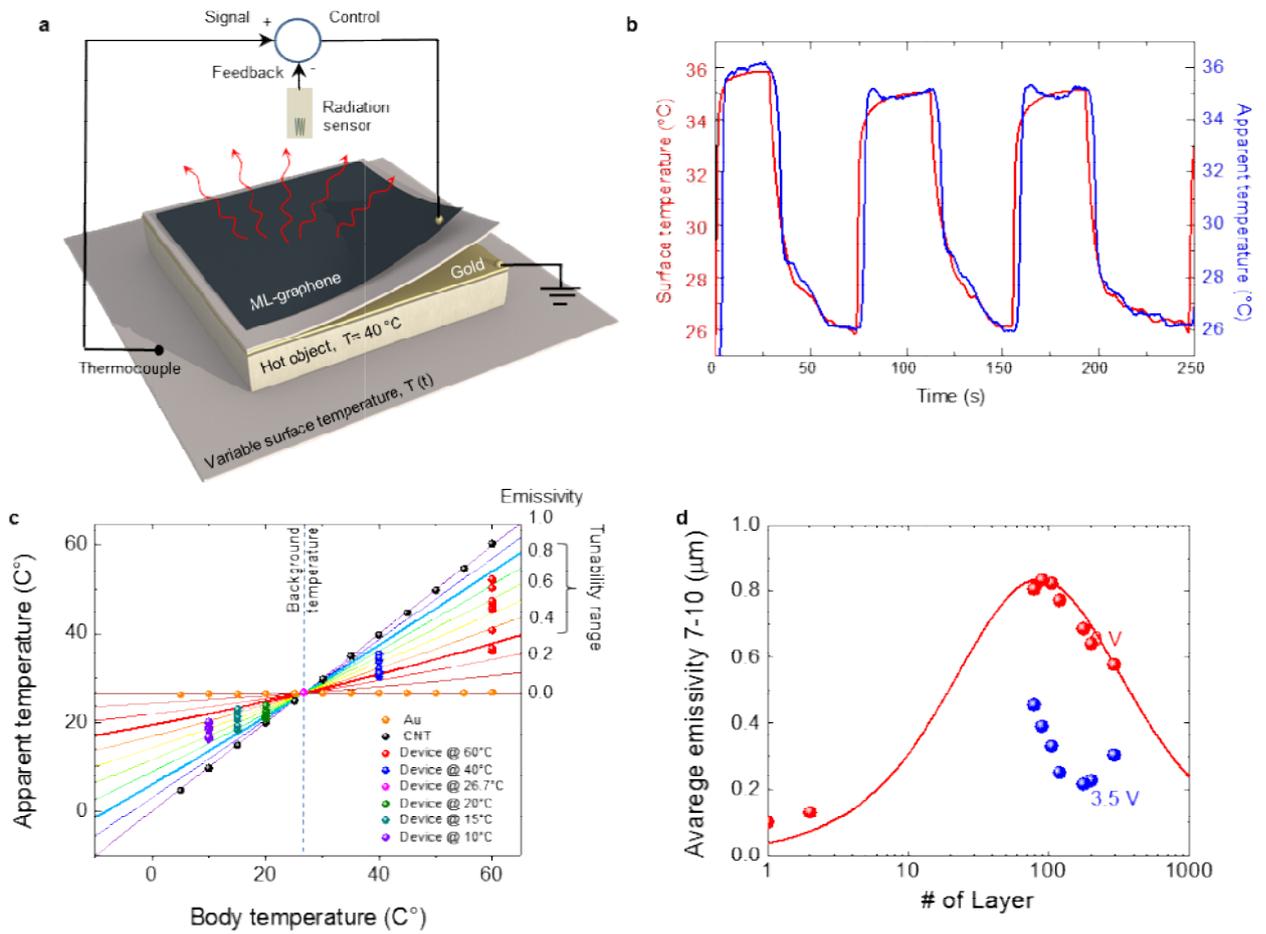

**Figure 4.** Adaptive thermal camouflage systems. (a) Schematic drawing of the device capable of blending its thermal appearance into a variable background temperature. (b) Time trace of the surface temperature and the apparent temperature of the device. (c) Apparent temperature of a surface plotted against the actual body temperature with different emissivity. The lines show the calculations and the scattered dots represent the measured data. (d) Layer dependence of the averaged emissivity of multilayer graphene (between 7-14 μm wavelengths for the device configuration given in Figure 1a). The maximum emissivity of 0.8 is obtained around 100 layers. The scattered plot shows the measured values at 0 and 3.5V bias voltages.

Finally, we would like to demonstrate an integration scheme which yields more complex reconfigurable thermal images. Figure 5a shows the multipixel device consisting of large area continuous graphene film on PE substrate and 5x5 arrays of individually addressable gold electrodes deposited on a printed circuit board. In this layout, the graphene film is wired to the ground electrode. By controlling voltage of a pixel with an external circuit, we were able to confine the intercalation within the pixel thus results modulation of local emissivity. Figure 5b shows three thermal images of the device with different voltage configurations. For low and high emissivity, we applied -3.5 and 0 V to the pixels, respectively. A temperature contrast of 10 °C can be obtained at each pixel individually (Figure 5c) and can be switched in 0.1 sec (Figure 5d). The crosstalk between the pixels is negligible. With this area selective intercalation, we generated complex thermal images such as a text "HELLO" (See Movie 5). The size of the pixels can be scaled down to millimeter without a significant crosstalk. These devices can also be fabricated by pattering the graphene layer and using different addressing mechanisms (see Figures S13 and S14). These results show that, our approach can be used to disguise the shape and temperature of objects in thermal imaging systems. Furthermore these devices can also operate as adaptive IR-mirrors.

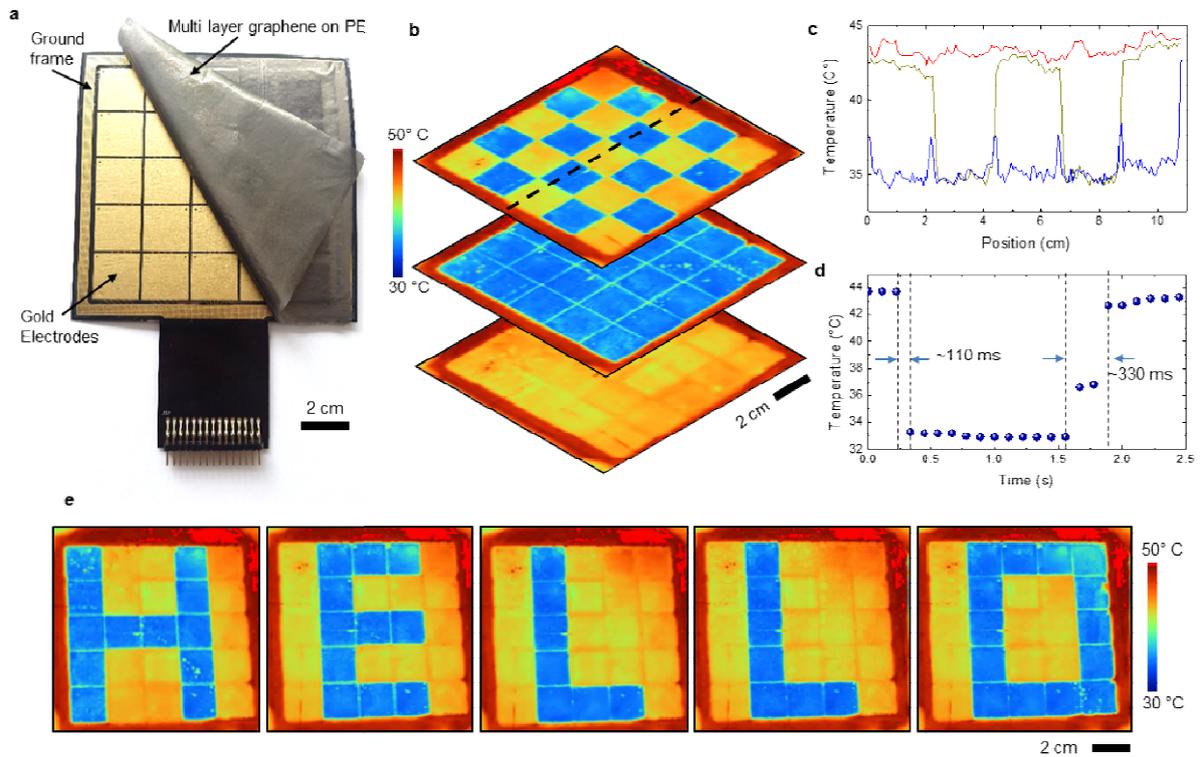

**Figure 5.** Multipixel active thermal surface. (a) Photograph of the device consisting of 5x5 arrays of individually addressable pixels with an area of 2x2 cm$^2$. The pixels are defined by the pattered gold electrodes on a printed circuit board and the top graphene layer is wired to the ground electrode. (b) Thermal camera images of the device (heated to 55°C) for three different voltage configurations; all pixels are grounded (bottom), all pixels are biased to -3.5 V (middle) and pixels are biased alternatively between 0 and -3.5V. (c) Line profile of the apparent temperature of the device shown in (b). (d) Time-trace of the apparent temperature of the device switched between different voltage configurations. (e) Complex thermal images of text "HELLO" generated by the device.

In conclusion, we have developed a new class of active thermal surfaces capable of efficient real-time electrical-control of their thermal emission over the full infrared spectrum. We showed that emissivity of multilayer graphene electrodes can be controlled electrically between 0.8 down to 0.3 with a bias voltage less than 4 V. Using these active surfaces, we have demonstrated adaptive camouflage systems that can disguise hot surfaces as cold and cold ones as hot in a thermal imaging system. Simplicity of the layered device structure together with the efficient modulation over broad IR spectrum (from 2 to 25 µm) provides an unprecedented ability for adaptive thermal camouflage. These active surfaces are flexible which enable their integration with nonplanar surfaces, such as soft robotic systems[2]. Furthermore, these devices can operate at high temperatures and under high vacuum conditions due to low vapor pressure of the ionic liquids enabling us to monitor the intercalation process using X-ray photoelectron spectroscopy. Our results provide a significant step for realization of adaptive thermal management, which could enable new technologies, not only for thermal camouflage but also for adaptive IR optics and adaptive heat shields for satellites[23].

**Methods**:

**Synthesis and transfer printing of multilayer graphene:** We synthesized multilayer graphene on 50μm thick Ni foil substrates (*Alfa Aesar Item #12722*) using a chemical vapour deposition system. By adjusting the growth temperature between 900 $^0$C, to 1050 $^0$C, we controlled the number of graphene layers from 60 to 100 layers. During the growth, we used 30 sccm of $CH_4$ and 100 sccm Ar and 100 sccm $H_2$ gases at ambient pressure. The growth duration was 5 minutes. After cooling the samples to room temperature, we etched the Ni foil in a $FeCl_3$ solution (1 M). We transferred the ML-graphene on a clean water surface. The surface of graphene is hydrophobic, which allows free standing ML-graphene film on the surface of water. By immersing the polyethylene membrane into the water, graphene conformally coat the surface.

**Fabrication of active thermal surfaces:** After the transfer process, we injected room temperature ionic liquid electrolyte [DEME][TFSI] (98.5%, Diethylmethyl (2methoxyethyl)ammoniumbis(trifluoromethylsulfonyl)imide, Sigma-Aldrich, 727679) into the membrane and attached copper wires on the ML-graphene with a conductive tape. To fabricate the gold electrode, we evaporated 5 nm Ti adhesive layer and 100 nm Au on 25 μm thick heat resistive nylon using thermal evaporation. We placed the PE membrane on the gold coated nylon.

**Thermal imaging**: The thermographs of the samples were recorded using FLIR A40 thermal camera. The camera renders the thermographs using constant emissivity of 1.

**Electrical measurements**: To apply the bias voltage to the devices, we used Keithley 2400 source measure unit. We recorded both voltage and charging current during the intercalation and de-intercalation process. To measure sheet resistance, we used 4-point resistance measurement system (Nano Magnetics Inc.) which includes two separate source meters (Keithley 2400 and 2600). The first power supply applies the bias voltage between the ML-graphene and the gold electrodes to initiate intercalation and the second one measures the sheet resistance.

**Spectroscopic characterization**: Thermal emission measurements were performed using Bruker Vertex 70v Fourier transform infrared spectrometer (FTIR). The devices were placed on a hot plate at constant temperature of 55 °C. The hot plate is aligned to the emission port of the spectrometer. We used wide range DLATGS detector (D201/BD) and wide-range beam splitter (T240) in the spectrometer. A Thermo Fisher K-Alpha spectrometer was used for XPS characterizations.

**Acknowledgements**

C.K. acknowledges the financial support from European Research Counsel for ERC-Consolidator Grant SmartGraphene. 682723. C.K. acknowledges BAGEP Award of the Science Academy.

**Author contributions**: C.K. and O.S. proposed the idea and planned the experiments. B.U synthesized the samples. B.U and O.S. fabricated the devices. B.U., O.Y., O.S. and C.K. performed the experiments. S.S. performed the XPS measurements and analysed the data. S.B., S.A., N.K., O.B. and O.S. helped for the measurements and electromagnetic modelling of the devices. S.O. designed the multipixel device. C.K. analysed the data and wrote the manuscript. All authors discussed the results and contributed to the scientific interpretation as well as to the writing of the manuscript.

**Additional information**: Authors declare no competing financial interests.